\documentclass[prb,twocolumn,,amsmath,amssymb,floatfix]{revtex4}
\usepackage{graphicx}
\usepackage{bm}

\newcommand{\be}{\begin{equation}}
\newcommand{\ee}{\end{equation}}

\def \be{\begin{equation}}
\def \ee{\end{equation}}
\def \ba{\begin{array}}
\def \ea{\end{array}}
\def \bea{\begin{eqnarray}}
\def \eea{\end{eqnarray}}

\def \W{{\Omega}}

\def \a{{\alpha}}
\def \t{{\theta}}

\def \nd{{^{\vphantom{\dagger}}}}
\def \yd{^\dagger}
\def \av#1{{\langle#1\rangle}}

\begin{document}

\title{Interference between independent fluctuating condensates}

\author{ Anatoli Polkovnikov$^{1,2}$, Ehud Altman$^{1,3}$,
and Eugene Demler$^1$}
\affiliation
{$^1$Department of Physics, Harvard University, Cambridge, MA 02138\\
$^2$ Department of Physics, Boston University, Boston, MA 02215\\
$^3$ Department of Condensed Matter Physics, The Weizmann
Institute of Science Rehovot, 76100, Israel}

\date{\today}

\begin{abstract}
We consider a problem of interference between two independent
condensates, which lack true long range order. We show that their
interference pattern contains information about correlation
functions within each condensate. As an example we analyze the
interference between a pair of one dimensional interacting Bose
liquids. We find universal scaling of the average fringe contrast
with system size and temperature that depends only on the Luttinger
parameter. Moreover the full distribution of the fringe contrast,
which is also equivalent to the full counting statistics of the
interfering atoms, changes with interaction strength and lends
information on high order correlation functions. We also demonstrate
that the interference between two-dimensional condensates at finite
temperature can be used as a direct probe of the Kosterlitz-Thouless
transition. Finally, we discuss generalization of our results to
describe the intereference of a periodic array of independent
fluctuating condensates.
\end{abstract}

\maketitle

\section{Introduction}
An important property of Bose Einstein condensates (BEC) is the
existence of a coherent macroscopic phase. Thus, a crucial benchmark
in the study of such systems was the observation of interference
fringes when two independent condensates were let to expand and
overlap \cite{ketterle}. This ``two slit'' experiment was carried
out with cold atoms in three dimensional harmonic traps, where a
true condensate exists. The interference fringe amplitude should
then be proportional to the condensate fraction, as was indeed
observed. However, with current trapping technology it is possible
to confine the bosonic atoms to one~\cite{paredes,weiss,fertig} or
two dimensions~\cite{stock} where a true condensate may not exist.
Instead, these systems are characterized by off-diagonal
correlations that decay as a power-law or exponentially in space.
What is the interference pattern that arises when two such imperfect
condensates are let to expand and overlap? This question is not just
of general academic interest. Recently there have been a number of
beautiful experiments where the interference between independent
condensates was directly observed, see for example
Refs.~[\onlinecite{zoran, jorg, ketterle1}].

Here we address this problem theoretically and show that the result
depends crucially on the correlations within each condensate.
Therefore, such an experiment would provide a direct and simple
probe of the spatial phase correlations. In principle spatial phase
correlations may also be extracted from juggling experiments
\cite{hagley,bloch2000,petrov2000} or the momentum distribution
measured by the free expansion of a single condensate\cite{paredes}.
However, creating strongly interacting low dimensional systems
typically requires using low density atomic gases, which makes
juggling experiments very challenging.  Also the highly anisotropic
expansion of low dimensional condensates inhibit measurements of the
momentum distribution in the slowly expanding longitudinal
direction. A method for probing the phase correlations directly in
real space would therefore be very useful.

\section{results and disscussion}
\begin{figure}[ht]
\includegraphics[width=8cm]{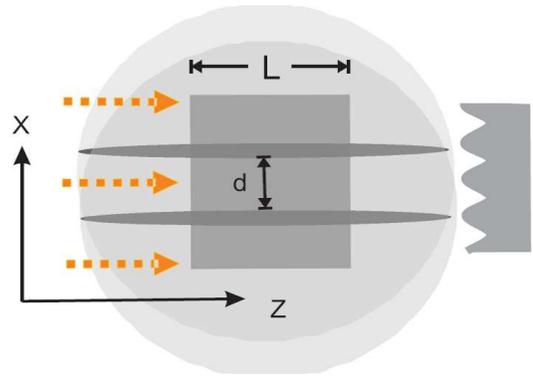}
\caption{Schematic view of the possible experimental setup, which
produces interference pattern between two independent 1D
condensates. Here $L$ and $d$ are the imaging length and the
separation between the condensates, respectively.}
\label{fig:tubes}
\end{figure}

\begin{figure}[ht]
\includegraphics[width=8cm]{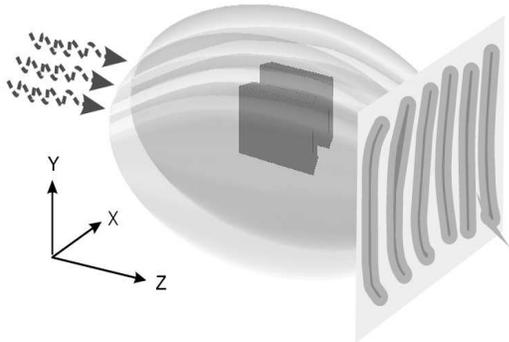}
\caption{Same as in Fig~\ref{fig:tubes} but for two dimensional
condensates.}
\label{fig:2d}
\end{figure}

The simplest geometry that we consider is illustrated in
Fig.~\ref{fig:tubes}. It consists of two parallel one dimensional
(1D) condensates a distance $d$ apart. After the atoms are released
from the trap they are let to expand to a transverse size much
larger than $d$, while no significant expansion occurs in the axial
direction.
An absorption image is then taken by a probe beam
directed along the condensate axis. A similar setup is considered
for two dimensional (2D) condensates on parallel planes (see Fig.
\ref{fig:2d}).
As usual, the absorption image gives the instantaneous three
dimensional density profile, integrated along the beam axis. $
{\rho}(x)=\int_0^L dz \,a_{tof}^\dagger (x,z)\, a_{tof}(x,z) $,
where $a_{tof}^\dagger$ are the Bose creation operators with the
subscript ``$tof$'',emphasizing that the corresponding operators
are taken after free expansion of atoms, $z$ is the axial coordinate
and $x$ is the coordinate along the detector (see
Fig.~\ref{fig:tubes}). The length $L$ is typically given by the
focal length of the imaging beam. It may also be controlled more
precisely by applying magnetic field gradients, so that only a
specified section of the cloud is resonant with the probe light. In
principle, one can consider an experiment with a probe beam
orthogonal to the plane containing two parallel $1D$ condensates. In
this case, it is possible to integrate the resulting interference
image within an arbitrary interval and obtain dependence of the
interference contrast on $L$ (note that this dependence
characterizes a single run and a series of experiments is still
needed to find the average contrast).  The other advantage of this
set-up is that it can reveal the presence of dipolar oscillations in
individual condensates. These modes correspond to the center of mass
motion and are not affected by interactions. Dipolar oscillations
induce an overall tilt in the interference peak position and can be
easily removed by integrating $\rho(x)$ along a line tilted with
respect to the $z$ axis. However, since most of the current
experimental systems do not allow imaging beams that are
perpendicular to 1D condensates, we concentrate on the setup shown
in Fig.~\ref{fig:tubes}.

To discuss the interference contrast we consider the correlation
function of the density operator
\begin{eqnarray}
&&\langle \rho(x_1) \rho(x_2) \rangle = \delta(x_1-x_2) \int_0^L dz
\rho(z)\nonumber\\
&& +\int_0^L d z_1 d z_2 \langle a_{tof}^\dagger(r_1)
a_{tof}^\dagger(r_2) a_{tof}(r_1) a_{tof}(r_2) \rangle,
\label{interference}
\end{eqnarray}
where $r_i$ stands for $(x_i,z_i)$. Single particle operators in
(\ref{interference}) should be taken after the expansion time $t$.
We can relate them to operators before the expansion:~\cite{roth} $
a_{tof}(x,z)=a_1(z) e^{i Q_1(x) x - i Q_1^2 t/2m} +a_2(z) e^{i
Q_2(x) x - i Q_1^2 t/2m} $, with $a_{1,2}$ being operators in the
two condensates and $Q_{12}=m(x \pm d/2)/\hbar t$. We thus find that
the correlation function in (\ref{interference}) has an oscillating
component at wavector $Q=md/\hbar t$
\begin{eqnarray}
\langle \rho(x_1) \rho(x_2) \rangle_{\rm int} = \langle |A_Q|^2
\rangle \left[e^{i Q(x_1-x_2)} + {\rm c.c.} \right],
\label{AQ2a}\\
\langle |A_Q|^2 \rangle =  \int_0^L dz_1 dz_2 \langle
a_2^\dagger(z_1) a_1(z_1) a_1^\dagger(z_2) a_2(z_2) \rangle.
\label{AQ2}
\end{eqnarray}
Here $A_Q=\int dz a\yd_1(z)a\nd_2(z)$ is the quantum observable
corresponding to the amplitude of the interference fringes. It can
be extracted from the time of flight (TOF) absorption image by
taking the Fourier transform of the density profile. Alternatively
one can directly probe $A_Q^2$ by studying the oscillating
component in the density autocorrelation function. Both methods
were successfully used in recent
experiments~\cite{marcus,bloch2005}. In practice it might be
easier to study the interference contrast rather than the absolute
value of the fringe amplitude. In this case one has to divide
$A_Q$ by the imaging length $L$. If the two condensates are
decoupled from each other, the expectation value of $\langle A_Q
\rangle $ vanishes. This does not mean that $|A_Q| $ is zero in
each individual measurement but rather shows that the phase of
$A_Q$ is random~\cite{dalibard}. Said differently, $A_Q$ is finite
in each experimental run, but its average over many experiments
vanishes. To determine the amplitude of interference fringes in
individual measurements one should consider an expectation value
of the quantity that does not involve the random phase of $A_Q$.
This naturally brings us to equation (\ref{AQ2}). From shot to
shot $|A_Q|^2$ fluctuates as well, and Eq.~(\ref{AQ2}) gives its
average value.

If the two condensates are identical (but still independent) we may
simplify expression~(\ref{AQ2}):
\be
\langle |A_Q|^2 \rangle = L\int_0^L dz \av{a\yd(z)a\nd(0)}^2.
\label{main}
\ee
Here we  neglected boundary effects by integrating over the center
of mass coordinate and assuming that the correlations depend only on
$(z_1-z_2)$. Eq.~(\ref{main}) can be generalized for the case of
parallel 2D condensates by taking $z$ to represent the planar
coordinates.

To gain intuition into the physical meaning of the fringe amplitude
let us first address two simple limiting cases. First, consider the
situation where $\langle a^\dagger (z) a(0)\rangle$ decays
exponentially with distance with a correlation length $\xi<<L$. Then
Eq. (\ref{main}) implies that $|A_Q|\propto \sqrt{L \xi}$, which has
a very simple physical meaning. Since the phase is only coherent
over a length $\xi$ the system is effectively equivalent to parallel
chains with $L/\xi$ pairs of independent condensates. Each pair
contributes interference fringes with a constant amplitude
proportional to $\xi$ and a random phase. The total amplitude $A_Q$
is therefore the result of adding $L/\xi$ independent vectors of
constant length $\xi$ and random direction, hence we get
$\sqrt{L\xi}$ scaling. Note that the interference contrast, which is
proportional to $A_Q$/L, the ratio of fringe amplitude to the
background signal, scales as $\sqrt{\xi/L}$. This observation is
similar in spirit to that made in Ref. [\onlinecite{zoran}] of
interference between 30 independent condensates in a chain. Fringes
can be seen, though their average amplitude is suppressed by a
factor of $\sqrt{30}$ compared to the interference between two
condensates. Now consider the opposite limit of perfect condensates,
for which $\av{a\yd(z)a\nd(0)}$ is constant. In this case Eq.
(\ref{main}) implies that $|A_Q| \propto L$. Pictorially this is the
result of adding vectors with a uniform phase, resulting in a fringe
amplitude which scales as the total size of the system. This is
essentially the result of the experiment in
Ref.~[\onlinecite{ketterle}].

\subsection{One dimensional Bose liquids}
We proceed to discuss the case of a
1D interacting gas. We first consider a system at sufficiently low
temperature, $\xi_T>>L$, where $\xi_T$ is the temperature dependent
correlation length defined in Eqs.~(\ref{thermal}) and
(\ref{thermal2}) below. In this regime the correlations decay as a
power law rather than exponentially. We therefore, expect that the
fringe amplitude will somehow interpolate between the two simple
limits considered above. Specifically, at long wavelengths the 1D
Bose gas is described by a Luttinger liquid~\cite{haldane} and the
long distance off-diagonal correlations behave as
\be
\langle a^\dagger(z) a(0)\rangle\sim \rho \left({\xi_h\over
z}\right)^{1/2K}.
\label{1dcor}
\ee
Here $\rho$ is the particle density, $\xi_h$ is the healing length,
which also serves as the short range cutoff, and $K$ is the
Luttinger parameter.  For bosons with a repulsive short-range
potential, $K$ ranges between $1$ and $\infty$, with $K=1$
corresponding to strong interactions, or ``impenetrable'' bosons
while $K\to \infty$ for non interacting bosons. Substituting
Eq.~(\ref{1dcor}) into Eq.~(\ref{main}) and assuming that $L\gg
\xi_h$ we arrive at one of our main results
\be
\langle |A_Q|^2 \rangle
= C\rho^2 L^2\left({\xi_h\over L}\right)^{1\over K},
\label{1dT0}
\ee
where $C$ is a constant of order unity. Thus we see that the amplitude
of the interference fringes ($\bar{A}_Q\equiv\sqrt{\av{|A_Q|^2}}$),
scales with a non trivial power of the imaging length.
In the non interacting limit ($K\to\infty$)
the scaling is linear $\bar{A}_Q\sim L$ as expected for a fully coherent
system. Interestingly $\bar{A}_Q\sim \sqrt{L}$ in the hard core limit ($K=1$),
as in systems with short range correlations which were discussed above.
A more careful examination of the integral shows that at
$K=1$ there are additional logarithmic corrections to the power law
scaling

\begin{figure}[ht]
\includegraphics[width=8cm]{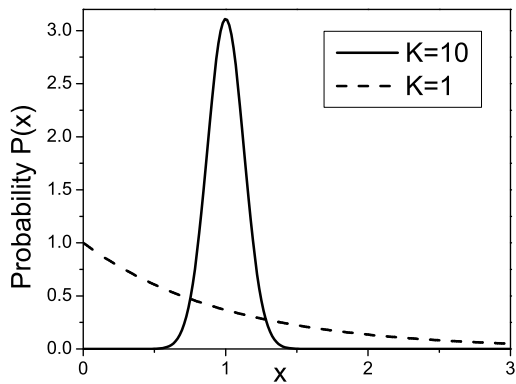}
\caption{ Schematic plots of the distribution function $P(x)$ of the
normalized intensity of interference fringes ($x=|A_Q|^2/\langle
|A_Q|^2\rangle)$ in strongly and weakly interacting regimes: $K\to
1$ and $K=10$.}
\label{fig2}
\end{figure}

Having derived the amplitude of interference fringes, an interesting
question is how this amplitude fluctuates from one experimental run
to the next. To answer this question one should consider higher
moments of the operator $|A_Q|^2$. We find that all moments have the
general form: $\av{|A_Q|^{2n}}=\av{|A_Q|^2}^n \mathcal{F}_n(K)$. It
follows that when $|A_Q|^{2}$ is normalized, its distribution
function $P(|A_Q|^{2}/\langle |A_Q|^2\rangle)$ is fully determined
by the Luttinger parameter $K$. In particular for large $K$ the
function $P(|A_Q|^{2}/\langle |A_Q|^2\rangle)$ becomes very narrow
characterized by the width $\sigma\equiv\sqrt{\langle
|A_Q|^4\rangle-\langle|A_Q|^2\rangle^2}/\langle|A_Q|^2\rangle\approx
\pi/(\sqrt{6}K)$. As expected, for $K\to\infty$ the width of the
distribution goes to zero\cite{Javanainen}.
In the opposite limit of $K\to 1$ the distribution takes another
simple limiting form: $P(x)\approx e^{-x}$. It is interesting to
point out that not only the scaling with $L$ but the whole
distribution of the interference fringes at $K\to 1$ is equivalent
to the one that arises in systems with short range correlations. In
Fig.~\ref{fig2} we plot schematically the two distributions obtained
in the opposite limits of strong and weak interactions. Note that
the true distribution at large $K$ is slightly asymmetric. We
address this issue in a separate paper~\cite{adgp}. We see that
there is an interesting crossover from a narrow distribution in
weakly interacting systems ($K\gg 1$) to a wide poissonian
distribution in the hard-core limit ($K\to 1$). In fact, one can
show that there is a close connection between the distribution
function $P(x)$ and the partition function of the boundary
Sine-Gordon problem. This gives a direct link between interference
of two 1D quasi-condensates and thermodynamics of a single impurity
problem in a Luttinger liquid. We defer the details of this problem
and the crossover from weak to strong interactions for future
investigation.~\cite{adgp}.

We showed above that it is possible to extract $K$ by analyzing
either the scaling of the fringe
amplitude with system size or by analyzing its distribution
at a given system size.
Another approach involves changing the angle $\t$ between
the probe beam and the condensate axis while keeping the imaging
length fixed. The resulting absorption image then corresponds to
integration of the cloud density along a line at an angle $\t$ to
the $z$ axis. Then Eq. (\ref{interference}) should be changed
accordingly. Let ${\bar x}$ be the transverse coordinate on the
screen ($z=0$), then in the second term in the right hand side of
(\ref{interference}) we substitute $x={\bar x}-z\tan\t$. On squaring
we obtain the analogue of Eq. (\ref{main})
\be
\av{|A_Q|^2(\theta)}=L\int_{0}^{L} dz \langle a^\dagger(z)
a(0)\rangle^2\mathrm \cos\left[q(\t)z\right],
\label{fourier}
\ee
where $q(\t)=k_0\tan\t$. For sufficiently large imaging
length ($qL>>1$) this yields
\be
\av{|A_Q|^2(\theta)}\sim (\rho^2 L\xi_h)(\xi_h q)^{1/K-1},
\label{Iq}
\ee
Thus the Luttinger parameter may be extracted from the angle
dependence of the fringe amplitude. For $qL<<1$ (i.e. very small
angles) Eq.~(\ref{fourier}) reduces to Eq.~(\ref{main}). Note that
if one uses the imaging beam orthogonal to the condensates, then
$\theta$ will be simply the angle between the $z$ axis and the
direction of integration of the interference contrast.

Before concluding this section let us address the effect of
temperature. It is well known that at any finite temperature the
off-diagonal correlations in a one dimensional Bose system must be
short ranged. Specifically at sufficiently long distances they decay
exponentially with a correlation length $\xi_T\sim 1/T$. The zero
temperature results presented above are valid at sufficiently low
temperature that $\xi_T>>L$. At higher temperature such that
$\xi_T<<L$, the scaling of the fringe amplitude with length must be
$|\bar{A}_Q|\sim\sqrt{L}$ as discussed above for systems with short
range correlations. However, as long as $\xi_T>>\xi_h$ the
temperature dependence of the fringe amplitude is universal and it
depends only on the Luttinger parameter:
\be
\langle |A_Q|^2 \sim L\rho^2\xi_h \left({\hbar^2\over m\xi^2}{1\over T}\right)
^{1-{1\over K}}.
\label{1dT}
\ee
This provides another experimental method to extract Luttinger physics.

\subsection{Two dimensional systems}
We now consider a pair of parallel 2D
condensates. In direct analogy to the 1D condensates, the imaging
axis may be taken parallel or at some angle to the plane of the
condensates. In the former case one should consider the scaling of
the fringe amplitude with imaging length, while in the latter case
the variation with angle.

It is well known that in two dimensions, long range order may exist
only at zero temperature. At sufficiently low temperatures
off-diagonal correlations are algebraic, with
\be
\langle a^\dagger({\bf r}) a({\bf 0})\rangle \sim\rho
{\left(\xi_h\over r\right)}^\alpha.
\label{2dcor}
\ee
for $r>>\xi_h$. On the other hand, above the Kosterlitz-Thouless
(KT) transition at $T=T_c$, the correlations decay exponentially. We
will show that this transition is characterized by a jump in the
behavior of the fringe amplitude, related to the
well known universal jump of the superfluid stiffness at $T_c$.

The exponent in (\ref{2dcor}) is given by
$\alpha=mT/2\pi\rho_s(T)\hbar^2$. For weakly interacting bosons at
temperatures well below $T_c$, $\rho_s(T)$ is simply equal to the
density $\rho$. As one approaches the transition $\rho_s$ is
renormalized by fluctuations, and at the transition
$\rho_s(T_c)=2mT_c/\pi\hbar^2$. Therefore the exponent $\a$ assumes
a universal value $\a_c=1/4$ at the transition. Thus for
temperatures $T<T_c$ we have $0<\alpha<\alpha_c$.

Let us now discuss the consequences of this physics to the
experimentally measurable fringe amplitude. As illustrated in Fig.
\ref{fig:2d},  the interference pattern is now truly 2D in the sense
that cuts along $x$ at different coordinate $y$ display a different
fringe pattern. To obtain a 1D pattern as a function of $x$ alone we
may integrate the image intensity over an ``integration length''
$L_y$. Recall that in addition the imaging process automatically
integrates over an imaging length $L_z$ along the $z$ axis. Now the
generalization of Eq.~(\ref{main}) to the 2D case is straight
forward
\be
\langle |A_Q|^2\rangle\sim L_y L_z\int\limits_0^{L_y}
dy\int\limits_0^{L_z} dz\, \langle a^\dagger(y,z), a(0,0)\rangle^2.
\label{2dT}
\ee
For simplicity we assume that $L_y$ and $L_z$ are scaled
simultaneously as $L_y=L_z=\sqrt{\W}$, with $\W$ the imaging area.
Then using Eqs.~(\ref{2dcor}) and (\ref{2dT}) we find that for
$T<T_c$
\be
\av{|A_Q|^2} \sim \W^{2-\alpha}.
\label{AQ2b}
\ee
So below the transition the scaling of $|A_Q|$ with size ranges from
linear at $T=0$ to $\W^{0.875}$ at the KT transition ($\a_c=1/4$).
On the other hand, for $T>T_c$ the correlations decay exponentially
and $|A_Q|\sim \sqrt{\W}$. Hence we find a universal jump at $T=T_c$
in the power characterizing the size dependence of $|A_Q|$. One can
also consider a setup where only dependence on one length, say
$L_y$, is studied while the other one $L_z$ is fixed. Then if
$L_y\gg L_z$, Eq.~(\ref{AQ2b}) immediately generalizes to
$\av{|A_Q|^2}\sim L_y^{2-2\alpha}$. In this case the power jump in
$|A_Q|$ is bigger: from $0.75$ to $0.5$ as $T$ crosses $T_c$. This
jump is a direct signature of the $KT$ physics, to be contrasted
with the result for 1D condensates, where the scaling power with system
size interpolated smoothly between $1$ and $2$. It should be noted,
however, that a 1D system on an optical lattice, which undergoes a
Mott transition at $T=0$, would display a universal jump similar to
the 2D case discussed here. In the same way one can study the shape
of the distribution of the interference amplitude and find that as
$T$ increases to $T_c$ the distribution gradually broadens but
always remains relatively narrow. On the other hand, as $T$ becomes
larger than $T_c$ the distribution assumes a broad poissonian form.

The analysis for imaging the 2D condensates with a slanted probe
beam can be carried over from the 1D case. The scaling of the
interference contrast with $q=k_0\tan\t$, at constant imaging area,
is then $\av{|A_Q|^2}\sim 1/q^{2-2\alpha}$ below the KT transition,
and $\av{|A_Q|^2}\sim 1/(1+q^2\xi^2)^{3/2}$ above it. Again the
transition is characterized by a universal jump of the power at
small $q$. We emphasize that $\theta$ can be either the angle
between the beam and the $z$ axis (see Fig.~\ref{fig:2d}) or the
angle between the $y$ axis and the direction of integration. The
latter is preferable, because within a single experimental shot it
is possible to obtain the whole angular dependence of $A_Q^2$.

Regardless of the experimental approach of choice, the interference
between a parallel pair of independent 2D condensates can serve as a
direct probe of Kosterlitz-Thouless physics.  However a word of
caution is in order. The correlation length, which coincides with
the healing length at very low temperatures\cite{petrov}
($\xi_T\approx\xi_h=\hbar/\sqrt{m g \rho}$), diverges at the $KT$
transition as $\xi_T\propto\exp(b/\sqrt{T_c-T})$~\cite{chaikin}.
Therefore, with increasing temperature, one has to probe the system
at increasing distances $r>>\xi_h(T)$ (or $L^{-1}<<q<<1/\xi_h(T)$)
in order to measure the asymptotic form of the correlation function
(\ref{2dcor}). This might hinder accurate determination of the
universal jump.

\subsection{Discussion}
Before concluding this paper we would like to make a few additional
remarks. We considered a pair of interfering quasi-condensates,
however most of our arguments can be generalized to the case of
several independent condensates. Of particular interest is a
periodic array of tubes\cite{paredes,weiss,fertig} or pancakes
created by an optical potential\cite{oberthaler,stock,inguscio}. The
interference pattern in this case shows correlations at a set of
wavectors $Q_n=nQ$, where $n$ is an integer and $Q$ is determined by
the distance between neighboring condensates. The size and angle
dependence of the average interference amplitudes for each of these
wavevectors should have the same scaling properties as two
quasi-condensates. However, the distribution function of fringe
amplitudes will be different. In particular, in the limit of a large
number of condensates, the distribution function should become very
narrow. This follows immediately from the observation that in this
limit, higher order correlation functions in TOF images are
dominated by products of two point correlation function in different
condensates, so there should be no broadening associated with
Eq.~(\ref{AQ2n}) below.

Another point worth making regards the possibility of making analogous
experiments with Fermions. For example, one can consider an
interference of two independent 1D fermionic systems. One obvious
difference from the bosonic case will be the change of sign in the
correlation function (\ref{AQ2a}), reflecting different statistics of
the fermions (this corresponds to fermion antibunching).  More
importantly the correlation function decays as $1/|x|^{1/2(K+1/K)}$,
i.e. as $1/x$ or faster. This means that the integral in (\ref{main})
is dominated by short distances, where the Luttinger liquid
description is not sufficient, and that the integral converges as $ L
\rightarrow \infty$. Infrared convergence of (\ref{main}) implies
trivial scaling $|A_Q|\propto \sqrt{L}$. Moreover, the integrals
appearing in all moments of the distribution are similarly infrared
convergent, which results in a Poissonian fringe distribution at large
$L$: $P(x)\propto e^{-x}$, as found for bosons at high temperature. In
order to extract information on the Luttinger parameter one can
analyze the decay of density-density correlations in the noise
$\av{\rho_{int}(x,z_0)\rho_{int}(x,z_0+ z)}$ as a function of $ z$, which are
directly related to the integrand of (\ref{main}). We note that these
correlations have an oscillating component, similar to Friedel
oscillations, with wavevector $2k_f$. The oscillating component
appears as a peak (cusp singularity) in the angular dependence of the
interference contrast at an imaging angle $k_0\tan\t=2k_f$ (see
Eq.~(\ref{fourier})). The shape of this cusp as well as of the cusp at
$\theta=0$ holds information on the Luttinger parameter. A more
detailed analysis of the fermionic case lies beyond the scope of this
work and will be addressed in future work.

In conclusion, we analyzed the interference between two independent
quasi-condensates. We showed that scaling properties of
interference fringes directly probe the algebraic off-diagonal
correlations. In particular, for one dimensional condensates the
scaling with imaging length or with temperature allows to extract
the Luttinger parameter. In the case of two dimensional condensates
this method provides a unique probe of the Kosterlitz-Thouless
transition. We also argued that in the 1D case one can  use the
distribution function of the interference amplitude (which is also
equivalent to the full counting statistics of interfering bosons) as
the qualitative probe of the Luttinger constant. In particular, at
$K\gg 1$ the distribution is narrow and at $K\to 1$ or at finite
temperatures it becomes wide Poissonian (see Fig.~\ref{fig2}). In
the 2D case we expect a sharp change in the shape of the distribution
function at the KT transition. The scaling analysis remains intact
if more than two independent condensates are present, but the
distribution functions can no longer be used as a probe of the
correlations.

\section{methods}

\subsection{Luttinger liquid parameter}

The Luttinger liquid provides a universal long-wavelength
description of one dimensional interacting Bose liquids which allows
to calculate the long distance behavior of correlations such as
(\ref{1dcor}). In certain regimes it is possible to derive the
Luttinger parameter $K$ and the healing length $\xi_h$ from the
microscopic interactions. In particular, for bosons with {\em weak}
contact interactions, relevant for ultra cold atom systems discussed
in this work, one can use Bogoliubov theory to obtain~\cite{castin,
cazalilla}:
\be
\xi_h\approx {1\over \rho\sqrt{\gamma}},\; K\approx
{\pi\over\sqrt{\gamma}}\left(1-{\sqrt{\gamma\over
2\pi}}\right)^{-1/2}.
\ee
Here $\gamma\equiv 2/(a_{1d}\rho)\lesssim 1$ is a dimensionless
measure of the interaction strength and $a_{1d}$ is the one
dimensional scattering length. Analytic expressions for these
parameters may also be derived in the limit of strong contact
interaction $\gamma\gg 1$ ~\cite{cazalilla}:
\be
\xi\approx 1/\rho,\; K\approx 1+{4\over\gamma}.
\ee

\subsection{Moments of the fringe amplitude}
All the moments in the distribution of $|A_Q|^2$ can be obtained by generalizing the
two point correlation function in Eq. (\ref{AQ2}) to the $2n$-point
correlation function
\begin{eqnarray}
&&\langle |A_Q|^{2n} \rangle = \int_0^L\dots\int_0^L dz_1 \dots dz_n
dz_1' \dots dz_n'\nonumber\\
&&~~~~~~~~~~~~~~~~  | \langle a^\dagger(z_1) \dots a^\dagger(z_n)
a(z_1') \dots a(z_n') \rangle |^2.
\label{AQ2n}
\end{eqnarray}
For bosons with repulsive interactions described by the Luttinger parameter
$K$ we have
\begin{eqnarray}
&&\langle |A_Q|^{2n} \rangle = (\tilde C\rho^2
L^2)^n\left({\xi_h\over L}\right)^{n\over K}\int_0^1\dots\int_0^1 d
\omega_1 \dots
d\omega_n'\nonumber\\
&&~~~~~~~~~~~~~~~~ \left| \frac{\prod_{i<j} |\omega_i-\omega_j|
\prod_{i<j} |\omega_i'-\omega_j'|} {\prod_{ij}|\omega_i-\omega_j'|},
\right|^{1/K}\!\!\!\!.
\label{HigherMoments}
\end{eqnarray}
which is precisely of the form $\av{|A_Q|^{2n}}=\av{|A_Q|^2}^n
\mathcal{F}_n(K)$. Integrals of the type appearing in
Eq.~(\ref{HigherMoments}) have been discussed by Fendley
et.al.~\cite{saleur}, who demonstrated that they can be calculated
using special properties of Jack polynomials. From the knowledge of
all moments one can, in principle, construct the full distribution
of the interference fringes amplitude. In this paper we only discuss
the limits of weak ($K\gg 1$) and strong ($K\approx 1$) interactions.

\subsection{Finite temperature correlations in one dimension}

The finite temprature off-diagonal correlations are given by \cite{cazalilla}:
\be
\langle a^\dagger(z) a(0)\rangle\sim \rho\,\xi_h^{1/2K}
\left({\pi/\xi_T\over \sinh\pi z/\xi_T}\right)^{1/2K},
\label{thermal}
\ee
where the thermal correlation length $\xi_T$ is
\be
{\xi_T\over \xi_h}\approx \left({\hbar^2\over m\xi_h^2}\right){1\over T}.
\label{thermal2}
\ee
Eq. (\ref{thermal}) is valid for sufficiently low temperatures so
that $\xi_T\gg\xi_h$, or equivalently $T\ll\hbar^2/m\xi_h^2$. For
$z\ll \xi_T$ Eq. (\ref{thermal}) reduces to the zero temperature
correlation~(\ref{1dcor}). In the opposite limit $z\gg \xi_T$ the
correlation function (\ref{thermal}) may be approximated by
\be
\langle a^\dagger(z) a(0)\rangle\sim \rho\,\left({\xi_h\over
\xi_T}\right)^{1\over 2K} \mathrm e^{-\pi z/2K \xi_T}.
\label{high_T}
\ee
As we already noted, for sufficiently low temperatures when
$\xi_T>L$, the fringe amplitude may be found from Eq.~(\ref{1dT0}).
For $L\gg\xi_T$ equation~(\ref{high_T}) implies
\be
\langle |A_Q|^2 \rangle \sim L\rho^2\xi_h^{1/K}\xi_T^{1-{1\over
K}}.
\ee
Finally, substituting (\ref{thermal2}) for $\xi_T$ gives (\ref{1dT}).

We also note that the angular dependence of the fringe amplitude at finite temperature
is given by:
\be
\av{|A_Q|^2(\theta)}\sim \rho^2 L\xi_h^{1/K}\xi_T^{1-{1\over
K}}{K/\pi\over 1+K^2q^2\xi_T^2/\pi^2}.
\label{Iq1}
\ee
From this expression it is harder to extract $K$ directly because of
uncertainty in determination of $\xi_h$ and hence $\xi_T$.

\section{Acknowledgements}
We are grateful to P.~Fendley, M.~Greiner, V.~Gritsev,
Z.~Hadzibabic, M.~Lukin, M.~Oberthaller, M.~Oshikawa,
J.~Schmiedmayer, V.~Vuletic, D.~Weiss, and K.~Yang for useful
discussions. This work was partially supported by BSF and NSF grant
DMR-0132874 .

\end{document}